\documentstyle[12pt,epsfig,dina4p]{article}
\def\g2{GeV$^2$}
\def\q2{Q^2}
\def\oalphas3{O(\alpha_S^3)}

\def\Pma{I\!\!P}

\newcommand{\menor} {\mbox{\raisebox{-0.4ex}
{$\;\stackrel{<}{\scriptstyle \sim}\;$}}}

\begin{document}
\vspace{1 cm}
 
\begin{titlepage}
 
\title{{\bf Three-jet production in diffractive deep inelastic scattering at HERA}
\author{ ZEUS Collaboration}}
\date{}
 
\maketitle
 
\vspace{4 cm}

\begin{abstract}
\noindent
Three-jet production in the reaction $ep \rightarrow eXp$ has been studied with the ZEUS detector at HERA
 using an integrated luminosity of 42.74 pb${}^{-1}$.  The data were measured in the
kinematic region $5 < Q^2 < 100$ GeV$^2$, $200 < W < 250$ GeV and $23 < M_X < 40$ GeV. 
The diffractive signal was selected by requiring a large
rapidity gap in the outgoing proton direction. Jets were reconstructed
in the centre-of-mass system of $X$ 
using the exclusive $k_T$-algorithm. 
A sample of three-jet events in diffraction has been identified. Differential cross 
sections were measured  
as a function of the jet pseudorapidity and jet transverse momentum with 
respect to the virtual photon-Pomeron axis. The jets going in the Pomeron direction are broader than
those going in the virtual-photon direction. This is consistent with  models 
predicting that gluons are predominantly produced in the Pomeron direction and quarks in 
the virtual-photon direction.

\end{abstract}

\thispagestyle{empty}
\end{titlepage}
%
%
%
%
\evensidemargin-0.3cm                                                                              
\oddsidemargin-0.3cm                                                                               
\textwidth 16.cm                                                                                   
\textheight 680pt                                                                                  
\parindent0.cm                                                                                     
\parskip0.3cm plus0.05cm minus0.05cm                                                               
\def\3{\ss}                                                                                        
\pagenumbering{roman}                                                                              
                                                   %
\begin{center}                                                                                     
{                      \Large  The ZEUS Collaboration              }                               
\end{center}                                                                                       
  S.~Chekanov,                                                                                     
  M.~Derrick,                                                                                      
  D.~Krakauer,                                                                                     
  S.~Magill,                                                                                       
  B.~Musgrave,                                                                                     
  A.~Pellegrino,                                                                                   
  J.~Repond,                                                                                       
  R.~Yoshida\\                                                                                     
 {\it Argonne National Laboratory, Argonne, IL, USA}~$^{n}$                                        
\par \filbreak                                                                                     
  M.C.K.~Mattingly \\                                                                              
 {\it Andrews University, Berrien Springs, MI, USA}                                                
\par \filbreak                                                                                     
  P.~Antonioli,                                                                                    
  G.~Bari,                                                                                         
  M.~Basile,                                                                                       
  L.~Bellagamba,                                                                                   
  D.~Boscherini$^{   1}$,                                                                          
  A.~Bruni,                                                                                        
  G.~Bruni,                                                                                        
  G.~Cara~Romeo,                                                                                   
  L.~Cifarelli$^{   2}$,                                                                           
  F.~Cindolo,                                                                                      
  A.~Contin,                                                                                       
  M.~Corradi,                                                                                      
  S.~De~Pasquale,                                                                                  
  P.~Giusti,                                                                                       
  G.~Iacobucci,                                                                                    
  G.~Levi,                                                                                         
  A.~Margotti,                                                                                     
  T.~Massam,                                                                                       
  R.~Nania,                                                                                        
  F.~Palmonari,                                                                                    
  A.~Pesci,                                                                                        
  G.~Sartorelli,                                                                                   
  A.~Zichichi  \\                                                                                  
  {\it University and INFN Bologna, Bologna, Italy}~$^{e}$
% ...
\newpage                                                                                           
$^{\    1}$ now visiting scientist at DESY \\                                                      
$^{\    2}$ now at Univ. of Salerno and INFN Napoli, Italy \\                                      
% ...
                                                           %
                                                           %
\newpage   
                                                           %
                                                           %
\begin{tabular}[h]{rp{14cm}}                                                                       
$^{a}$ &  supported by the Natural Sciences and Engineering Research                               
          Council of Canada (NSERC)  \\                                                            
$^{b}$ &  supported by the German Federal Ministry for Education and                               
          Science, Research and Technology (BMBF), under contract                                  
          numbers 057BN19P, 057FR19P, 057HH19P, 057HH29P, 057SI75I \\                              
$^{c}$ &  supported by the MINERVA Gesellschaft f\"ur Forschung GmbH, the                          
          Israel Science Foundation, the U.S.-Israel Binational Science                            
          Foundation, the Israel Ministry of Science and the Benozyio Center                       
          for High Energy Physics\\                                                                
$^{d}$ &  supported by the German-Israeli Foundation, the Israel Science                           
          Foundation, and by the Israel Ministry of Science \\                                     
$^{e}$ &  supported by the Italian National Institute for Nuclear Physics                          
          (INFN) \\                                                                                
$^{f}$ &  supported by the Japanese Ministry of Education, Science and                             
          Culture (the Monbusho) and its grants for Scientific Research \\                         
$^{g}$ &  supported by the Korean Ministry of Education and Korea Science                          
          and Engineering Foundation  \\                                                           
$^{h}$ &  supported by the Netherlands Foundation for Research on                                  
          Matter (FOM) \\                                                                          
$^{i}$ &  supported by the Polish State Committee for Scientific Research,                         
          grant no. 2P03B04616, 620/E-77/SPUB-M/DESY/P-03/DZ 247/2000 and                          
          112/E-356/SPUB-M/DESY/P-03/DZ 3001/2000\\                                                
$^{j}$ &  partially supported by the German Federal Ministry for                                   
          Education and Science, Research and Technology (BMBF)  \\                                
$^{k}$ &  supported by the Fund for Fundamental Research of Russian Ministry                       
          for Science and Edu\-cation and by the German Federal Ministry for                       
          Education and Science, Research and Technology (BMBF) \\                                 
$^{l}$ &  supported by the Spanish Ministry of Education                                           
          and Science through funds provided by CICYT \\                                           
$^{m}$ &  supported by the Particle Physics and                                                    
          Astronomy Research Council, UK \\                                                        
$^{n}$ &  supported by the US Department of Energy \\                                              
$^{o}$ &  supported by the US National Science Foundation                                          
\end{tabular}                                                                                      
                                                           %
                                                           %

\newpage
\pagenumbering{arabic}
\setcounter{page}{1}
\parskip 0mm
\parindent 5mm
\topmargin-3.0cm         
\section{Introduction}

One of the most important results from the $ep$ collider HERA is the observation
that about $10 \%$ of deep inelastic scattering (DIS) events exhibit  
a large rapidity gap (LRG) between the direction of the proton beam 
and that of the first significant energy deposition in the detector~\cite{lrg}.
These LRG events, $ep \rightarrow e X N$,  
result predominantly from the diffractive dissociation of the virtual photon,
 $\gamma^* p \rightarrow X p$, and have been 
interpreted in terms of the exchange of a colour-singlet object
known as the Pomeron ($\Pma$),  introduced to describe 
the energy dependence of total cross sections in 
hadron-hadron scattering \cite{dl}. 

The HERA data \cite{h1f2d,zeusf2d} have been analysed in 
 terms of a diffractive structure function, $F_2^D$~\cite{ingelman},
defined in analogy with the proton structure function, $F_2$. The results are consistent
 with a ``resolved Pomeron'' model where $F_2^D$ factorises into a Regge-inspired 
Pomeron flux and a Pomeron structure function. The latter has been analysed \cite{h1f2d} 
using the DGLAP evolution equations. The extracted parton densities  
are dominated by gluons. 

Alternatively, a number of QCD models \cite{pQCD} has been proposed to describe
 diffraction in DIS. In these models, the virtual photon dissociates into a 
$q\overline{q}$ or $q\overline{q}g$ state that interacts with the proton
 by the exchange of a gluon ladder. The 
diffractive cross section can then be formulated in terms of  
contributions from $q\overline{q}$  and $q\overline{q}g$ final states. 
The  $q\overline{q}g$ component is expected to  dominate the 
diffractive cross section at high masses, $M_X$ \cite{ryskin,bekw,zeusmx}. 

An important feature of the three-parton final state in both classes of models is that, in 
the rest frame of the system $X$ ($\gamma^* \Pma$ centre-of-mass system for the resolved Pomeron model), 
the $q\overline{q}$ system tends to populate the virtual-photon hemisphere, while the gluon is emitted
in the opposite direction (the Pomeron direction in the resolved Pomeron model).
In the resolved Pomeron model, the gluon is the remnant of the Pomeron left after the 
boson-gluon fusion process has produced a $q\overline{q}$ pair; as a remnant, it follows the 
original Pomeron direction. In dipole models, to leading order, the gluon in the 
$q\overline{q}g$ fluctuation from the virtual photon either couples directly to the exchanged 
gluon ladder or emerges from it, depending on the reference frame. In this case also, therefore, 
it follows the direction of the gluon ladder ($\equiv$ Pomeron in resolved Pomeron model) and the 
$q\overline{q}$ pair carries on in the direction of the virtual photon.

Studies of the hadronic final state and jet production in diffractive interactions have 
been  presented by both H1~\cite{h1fs} and ZEUS~\cite{zeusfs}. 
In this paper, the jet-like structure of the hadronic final state is studied  
with the aim of isolating three-parton final states by selecting three-jet events.   
As  was already observed in $e^+ e^-$ experiments \cite{wu}, three-jet 
topologies can be clearly separated only for  
centre-of-mass energies above about 20 GeV. 
The measurements are  thus performed in a restricted phase-space region populated by
high-mass diffractive systems. Such systems are a small fraction of the total diffractive cross section.
They are separated from a substantial background of non-diffractive processes by requiring a large rapidity gap.  
A search is made for jets using an algorithm that allows the  study of jet configurations aligned with 
respect to the $\gamma^* \Pma$ axis and ensures good parton-hadron correlations. 
The properties of the three-jet final states are used to test 
the expectation that the gluon from the  dominant $q\overline{q}g$
contribution  is emitted in the Pomeron direction. Cross sections are presented
as a function of the jet pseudorapidity and jet
 transverse momentum with respect to the $\gamma^* \Pma$ axis.
In addition, the internal structure  of the jets is measured  for jets 
found in either the photon or the Pomeron directions, with the aim of distinguishing between 
quark- and gluon-initiated jets. The different measurements are
compared to  models of diffraction in DIS.  


\section{Experimental setup}
The data were collected with the ZEUS detector at HERA during 1998-2000 when 
HERA collided 27.6 GeV electrons or positrons\footnote{Hereafter ``positron'' 
is used to refer to both electron and positron beams. For the $Q^2$ range studied,  
$e^-p$ and $e^+p$ scattering are assumed to give identical results since 
contributions from $Z^{\rm 0}$ exchange are negligible.} on 920 GeV protons. 
A total of  $7.96 \pm 0.14$~pb${}^{-1}$ of 
$e^- p$ data and $34.78 \pm 0.87$~pb${}^{-1}$ of $e^+ p$ data was used.
A detailed description of the ZEUS detector 
 can be found elsewhere \cite{zeus1,zeus2}. A brief outline of the most relevant
 components for this analysis is given below.   

The ZEUS high-resolution uranium-scintillator calorimeter (CAL) \cite{cal} 
covers $99.7 \%$ of the total solid angle. It is divided  into three parts 
with respect to the polar angle\footnote{The ZEUS coordinate system is a 
right-handed Cartesian system, with the $Z$ axis pointing in the proton beam 
direction, referred to as the ``forward direction'', and the $X$ axis pointing 
left towards the centre of HERA. The coordinate origin is at the nominal interaction 
point. The pseudorapidity is defined as $\eta = - \rm{ln}(\rm tan \frac{\theta}{2})$, where 
the polar angle, $\theta$, is measured with respect to 
the proton beam direction.} as viewed from the nominal interaction point:
forward (FCAL), rear (RCAL) and barrel (BCAL) calorimeters. 
Each part is subdivided longitudinally into electromagnetic (EMC) and 
hadronic (HAC) sections. The CAL energy resolution, as measured in  test-beams, is 
$\sigma(E)/E = 0.18/\sqrt{E \ ({\rm GeV})}$ for positrons and 
$\sigma(E)/E = 0.35/\sqrt{E \ ({\rm GeV})}$ for hadrons.

Charged particles were tracked by the central tracking detector (CTD) \cite{ctd}, which 
operates in a magnetic field of 1.43 T provided by a thin superconducting coil.
 The CTD is a drift chamber consisting of 72 cylindrical layers, organised in 
 9 superlayers, covering the region $15^{\circ} < \theta < 164^{\circ}$. The transverse 
momentum resolution for full-length tracks is 
$\sigma(p_t)/p_t = 0.0058 p_t \oplus 0.0065 \oplus 0.0014/p_t$, with $p_t$ in GeV.

In 1998, a Forward Plug Calorimeter (FPC) \cite{fpc} was installed 
in the $20 \times 20$ cm${}^2$ beam hole of the FCAL 
with a small hole of radius $3.15$ cm in the 
centre of the FPC to accommodate the beam pipe. This  
increased the forward calorimetric coverage 
by about 1 unit of pseudorapidity to  
$-3.8 \leq \eta \leq 5$. The FPC consisted of a lead-scintillator 
sandwich calorimeter divided into  electromagnetic and hadronic sections 
which were read out separately by wavelength-shifting  fibres and 
photomultipliers. The EMC (HAC) sections were segmented into cells 
of $2.4 \times 2.4$ cm${}^2$ ($4.8 \times 4.8$ cm${}^2$) area and 
constituted a total of 76 read-out channels. The energy resolution was  
$\sigma_E/E = 0.41/\sqrt{E} \oplus 0.062$ and 
$\sigma_E/E = 0.65/\sqrt{E} \oplus 0.06$ for positrons and pions, respectively, where 
the energies are measured in units of GeV \cite{fpc}.

The small-angle rear tracking detector (SRTD) \cite{srtd} was used to measure
the position of those positrons scattered at a sufficiently small angle to strike it.
It consists of two planes of scintillator 
strips attached to the front face of the RCAL. The SRTD signals resolve single 
minimum-ionizing particles and provide a position resolution of 0.3 cm.
The position of those positrons falling outside the SRTD acceptance was measured in the CAL.

The luminosity was measured via the positron-proton 
bremsstrahlung process, $e p \rightarrow e \gamma p$, 
using a lead-scintillator calorimeter (LUMI) \cite{lumi}  
located at $Z = - 107$ m in the HERA tunnel.

%
\section{Kinematics}
The kinematics of inclusive deep inelastic scattering  
$e(k) \ p(P) \rightarrow e(k') + anything$ is described by any 
two of the following variables:
$$
Q^2 = -q^2 = -(k-k')^2, \ \ \  x = \frac{Q^2}{2P \cdot q}, \ \ \ \ y = \frac{P \cdot q}{P \cdot k}, 
\ \ \ \ W^2 = \frac{Q^2(1-x)}{x} + m_p^2 \ ,
$$
\noindent
where $k$ and $k'$ are, respectively, the four-momenta of the initial and 
final positrons, $P$ is the initial proton four-momentum, $y$ is the fraction of the 
energy transferred to the proton in its rest frame, $m_p$ is the proton mass 
and $W$ is the $\gamma^*p$ centre-of-mass energy.
 
For the description of the diffractive process $e p \rightarrow eXp $, in  addition to the 
invariant mass of the system $X$, $M_X$, two further variables are introduced:
$$
x_{\Pma} = \frac{M_X^2+Q^2}{W^2+Q^2}, \ \ \ \ \beta = \frac{Q^2}{Q^2 + M_X^2} .
$$
In resolved Pomeron models, where the interaction is described as the exchange of a 
particle-like Pomeron, $x_{\Pma}$ is the fraction of the proton 
momentum  carried by the Pomeron and $\beta$ is the fraction of its 
momentum carried by the parton within it that is probed by the virtual photon.
 

\section{Monte Carlo simulations}

Three Monte Carlo (MC) models for the diffractive deep inelastic 
process $e p \rightarrow eXp $ have been considered in order to describe the 
diffractive hadronic final state.

The {RAPGAP 2.08/06} MC generator \cite{rapgap} implements  the
Ingelman-Schlein factorisable model \cite{ingelman} for the Pomeron.
The matrix elements for the $O(\alpha)$ process ($eq \rightarrow e q$) and for 
the $O(\alpha \alpha_S)$ processes ($e q \rightarrow~e q g$, $e g \rightarrow eq\overline{q}$)
 are included. To avoid divergences in the matrix elements for the 
$O(\alpha \alpha_S)$ processes for massless quarks, a cut of $p_T^2 > 3$ GeV${}^2$ was 
applied, where $p_T$ is the  transverse momentum of any of the outgoing partons 
with respect to the photon direction in the centre-of-mass frame of the hard 
scattering. The relative contribution of each component is determined 
by the quark and gluon densities within the Pomeron, which evolve according to the 
DGLAP  equations. The H1 parameterisation \cite{h1f2d} for the
Pomeron parton densities with a `hard gluon' carrying $\geq 80 \%$ of the momentum
at $Q^2_0 = 3$ GeV${}^2$ was used.
Two different RAPGAP MC samples were generated, in which the 
higher-order QCD corrections were treated in two different ways: either using the  
colour-dipole model (CDM)~\cite{cdm} as implemented in ARIADNE 4.08 \cite{ariadne} or 
using parton showers (PS) as implemented in LEPTO 6.1 \cite{lepto}. The Lund 
string-fragmentation scheme \cite{lund} as implemented in JETSET~7.4~\cite{jetset} 
was used for hadronisation. First-order electroweak corrections are 
taken into account via the HERACLES 4.6 program \cite{heracles}. 

The RIDI 2.0 MC generator \cite{ridi} implements the diffractive dissociation of the virtual photon 
 into $q\overline{q}$ and $q\overline{q}g$ final states 
following the approach of Ryskin \cite{ryskin}, in the framework of 
the leading logarithmic approximation of pQCD. Contributions 
from transversely and longitudinally polarised photons are included.
The predicted cross section is 
proportional to the square of the gluon momentum density in the proton. 
The CTEQ4M parameterisation \cite{cteq4} for the parton densities  was used. 
Final-state parton showers and fragmentation were implemented using {JETSET 7.4}. 
First-order radiative corrections were also included.

The SATRAP MC generator \cite{satrap}  is based on the Golec-Biernat and W\"usthoff model
 \cite{satura}. In this model, the diffractive cross section 
is expressed  as the convolution  of the photon  wave-function, calculated 
for  $q\overline{q}$ and $q\overline{q}g$ colour dipoles  and including the contributions of both 
transversely and longitudinally polarised photons, with the dipole cross 
section for scattering on the proton. The parameters of the model were tuned to 
describe the total $\gamma^* p$ cross section as measured at HERA. Final-state 
parton showers and fragmentation were simulated via JETSET 7.4. First-order radiative 
corrections were taken into account via the HERACLES 4.6 program.

During the course of this analysis, it became clear that neither RIDI nor SATRAP gave
a reasonable description of the data. As discussed in Section 8, this was traced to inadequacies in the 
implementation of the modelling of higher-order QCD processes in these models.
Most notably, no initial-state parton cascades were included, and the 
final-state QCD radiation from the gluon in the dominant  $q\overline{q}g$ contribution 
was suppressed. A new implementation of higher-order QCD processes in SATRAP was carried
 out \cite{taro,jung}, in which the CDM approach was implemented in a similar fashion to that in 
RAPGAP. This model is referred to as SATRAP-CDM.

Processes where the proton dissociates into a system $N$, $ep \rightarrow eXN$, 
were generated using  the EPSOFT 2.0 MC generator \cite{epsoft}. This simulates
events based on the triple-Regge formalism \cite{triple} and an $M_N$
distribution as measured for diffractive dissociation in $pp$
scattering. Samples with different parameters for the simulation of
the system $N$ were considered.

Non-diffractive DIS background was generated including first-order QED radiative corrections 
using {LEPTO 6.5} interfaced to {HERACLES 4.5} via DJANGOH 1.1 \cite{django}. The CTEQ4M  set 
of proton parton densities was used. The colour-dipole model was used to simulate parton 
cascades and fragmentation was performed using JETSET 7.4.

The RAPGAP-CDM model provides a good description of most of the measured 
distributions and was used to  study the accuracy of the kinematic variables and 
jet reconstruction, the efficiency for selecting events and the  
corrections for detector and resolution effects. The detector 
simulation was based on the GEANT~3.13 program \cite{geant}.


\section{Event reconstruction and selection}

The variables $Q^2$ and $W$ were determined from the information of the scattered 
positron. The hadronic final-state system, $X$, resulting from the dissociation of the virtual photon,  
was reconstructed using ``energy flow objects'', hereafter denoted as EFOs~\cite{f2},  assumed to have 
the mass of the pion. These EFOs combine information from charged tracks, as measured
 in the CTD, and energy clusters measured in the CAL and FPC. 
In the kinematic region considered in this analysis,  the invariant mass $M_X$ was
reconstructed on average with a systematic shift of $-14 \%$ and a relative resolution of $10 \%$,   
while the hadronic three-vector was reconstructed 
with a resolution of about $1$ GeV for the $X$ and $Y$ components 
and $\sim 2$ GeV for the $Z$ component. The four-momentum of the virtual photon, {${q}$}, 
 was determined with no significant bias and with a relative resolution of about $10 \%$ for the 
$X$ and $Y$ components and $\sim 7 \%$ for the $Z$ and energy components.

The triggering and online 
event selections were similar to those used for the ZEUS measurement of the 
structure function $F_2$ \cite{f2}. The offline selection criteria, applied to 
select diffractive DIS events, required:
\begin{itemize}

\item a scattered positron candidate (identified \cite{sinistra} via its pattern of energy deposition in the 
calorimeter) with energy, $E'_e$, greater than 10 GeV and $y < 0.95$. For 
an incoming positron of energy $E_e$, 
 $y$ was estimated from the energy and  the polar angle, $\theta'_e$, of the scattered 
positron by:
$$
y = 1- \frac{E'_e}{2E_e} (1 - \rm {cos} \theta'_e)  \ . 
$$
The impact position of the scattered positron in the RCAL was required to be outside a
 box of 26 $\times$ 20~cm${}^2$ around the beam pipe  to ensure a fully contained 
positron candidate. In addition,  candidates  
within the CTD acceptance region, but  without a matched charged track reconstructed in the CTD,
 were rejected;

\item $40 < \delta < 70$ GeV, where 
$\delta = \sum_i E_i (1 - \rm{cos} \theta_i)$  and the sum runs over all 
calorimeter cells, i.e. including those belonging to the 
scattered positron. This cut, together  with 
the previous requirements on the positron 
candidate, eliminated background from photoproduction 
events and beam-gas interactions producing fake scattered positrons;

\item a reconstructed vertex with 
$|Z| < 50$ cm and at least two associated tracks;

\item the presence of a large rapidity gap 
in the outgoing proton direction, to select diffractive events. 
Events with $\eta^{\rm max} < 3.0$ were selected, where $\eta^{\rm max}$ is the 
pseudorapidity of the most-forward cluster with energy above 400 MeV, as measured in the CAL and FPC;

\item at least three reconstructed EFOs in the hadronic final 
state to ensure that the event plane could be determined.

\end{itemize}

The measurements were limited to the following kinematic range:

\begin{itemize}

\item $5 < Q^2 < 100$ GeV${}^2$;

\item $200 < W < 250$ GeV;

\item $23 < M_X < 40$ GeV, where $M_X$ was  corrected 
for energy losses (see Section 6.1); 

\item $x_{\Pma} < 0.025$. 
 
\end{itemize}

In this region, the 
diffractive hadronic final state $X$ is well contained in the central 
detector, thus minimising energy losses through the forward and rear 
beam-pipe holes. A good separation of the diffractive signal from the non-diffractive 
background was  achieved: after the  $\eta^{\rm max}$ requirement, the background was
typically around $3 \%$ and at most $10 \%$, as estimated from MC studies.  
After all cuts, 7175 events remained.

\section {Three-jet search}

A search for jets was performed in the centre-of-mass system of the observed hadronic
 final state, $X$, using the exclusive $k_T$-algorithm \cite{kt}.
All EFOs in the event were used as input to the jet-finding procedure. 
For each pair of particles (EFOs) $i$ and $j$ in an event, the quantity
$$
y_{ij} = \frac{2 \cdot {\rm min}(E^2_i,E^2_j)(1-cos \theta_{ij})}{M_X^2}
\label{eq:yij}
$$
was computed, where $E_i$ is the energy of the $i^{th}$ particle and $\theta_{ij}$ is 
the angle between particles $i$ and $j$.  
The pair of particles with the smallest value of $y_{ij}$ was replaced by a 
{\it{pseudo-particle}} or cluster. The four-momentum of the cluster was 
determined using the ``E-recombination scheme'':
$$
E^{\rm cluster} = E_i + E_j, \ \ \ \ \ \ \ P^{\rm cluster}_{k} = P^i_k + P^j_k \ \ \ (k=X,Y,Z). 
$$
\noindent 
This clustering procedure was repeated until 
all $y_{ij}$ values exceeded a given threshold, $y_{\rm cut}$, and 
all the remaining clusters were then labelled as jets. 
For this recombination scheme,  the jets  
become massive and the total invariant mass, $M_X$, coincides 
with the invariant mass of the three-jet system. As applied in the 
centre-of-mass system, this algorithm produces at least two jets in every event.
The same jet search procedure 
was applied to the final-state hadrons for simulated events.\\

\begin{figure}[h]
\epsfysize=21cm
\epsffile{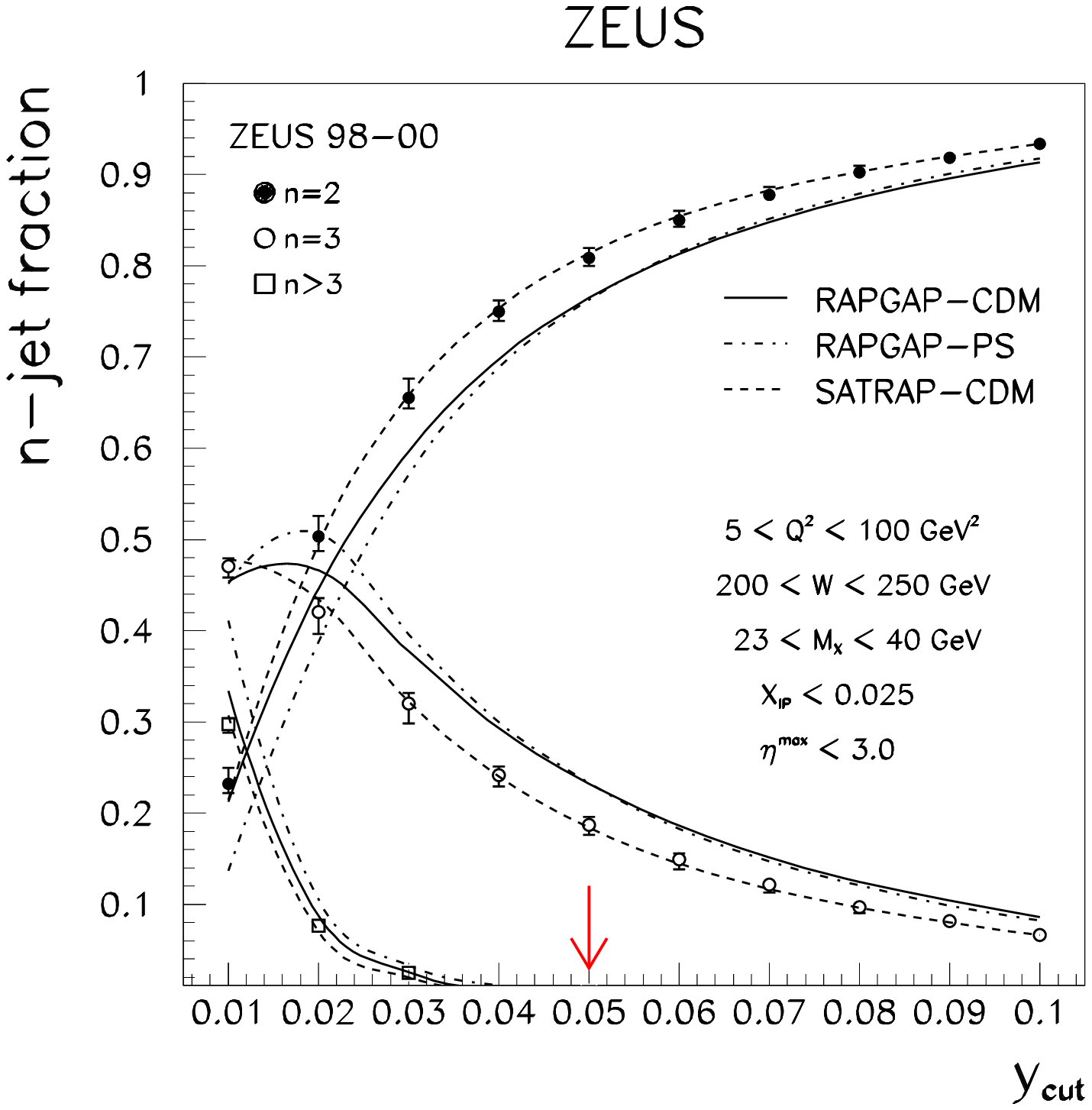}
\vspace{-2cm}
\caption{\label{fig1}{ 
Measurement of the $n_{\rm jet}$-fraction, for $n_{\rm jet}=2, 3$ and $ > 3$, 
as a function of the exclusive $k_T$-jet resolution parameter, $y_{\rm cut}$.  
The MC expectations 
from RAPGAP-CDM (solid lines), RAPGAP-PS (dashed-dotted lines) and SATRAP-CDM  (dashed lines)
are shown. The vertical arrow at $y_{\rm cut}=0.05$ indicates the value used in the subsequent analysis.
}
}
\end{figure}

Figure \ref{fig1} shows the measured $n_{\rm jet}$ fractions for  
$n_{\rm jet}=2,3$ and $>3$ as a function of the jet resolution parameter, $y_{\rm cut}$,   
in the region $0.01 < y_{\rm cut} < 0.1$.
The rate of three-jet production varies 
from $47 \%$ at $y_{\rm cut} = 0.01$ to  $8 \%$ at $y_{\rm cut} = 0.1$.
The measured jet fractions were compared with 
diffractive MC models. All the models show trends similar
to the data but only SATRAP-CDM provides a 
good description of the measurements. The 
predicted $n_{\rm jet}$ fractions are sensitive to the particular 
implementation of higher-order QCD corrections and 
hadronisation processes in each MC model. The differences 
observed between the predicted jet fractions from RAPGAP-CDM and 
RAPGAP-PS for $y_{\rm cut} \menor 0.03$ indicate where 
the dependence on the model used for the simulation of higher-order 
QCD contributions starts to be sizeable.\\ 

Good parton-hadron 
correlation in three-jet production was observed  in the MC simulation 
for $M_X > 20$ GeV. In this mass range, values for $y_{\rm cut}$ around 
$0.05$ gave the smallest hadronisation corrections.  
The parton-hadron jet correlation was studied using a MC sample with three 
jets found in the final state at  both parton and hadron levels. The angular 
distance between each parton and the corresponding jet of hadrons
was such that the partons were always well contained within 
the size of the hadron jets, which have a typical angular radius larger than 0.5 radians.

\subsection{Jet reconstruction and selection}
The reconstruction of the jet variables was studied using MC events with  
three jets at both hadron and detector levels.
Matched pairs of jets were selected and used to 
study the jet reconstruction in the detector. The characteristics of the reconstructed jet 
quantities were found to be as follows: 
\begin{itemize}
\item  the polar angle of the jet, $\theta^*_{\rm jet}$,
defined with respect to the 
$\gamma^* \Pma$ axis\footnote{In the centre-of-mass system, the Pomeron defines the 
positive $Z$ direction.}, was reconstructed 
with no significant systematic bias and a resolution   
of $\sim 6^{\circ}$ and $\sim 8^{\circ}$ for the two most energetic jets and the 
least energetic jet in the event, respectively;

\item the measured invariant mass, $M_X$, and jet energies, $E^*_{\rm jet}$, were corrected 
for the energy loss by a common factor of $1.14$, determined from MC studies. In the kinematic 
range considered, the jet energy  was reconstructed with a typical resolution of  
$10 \%$ and the jet energy correction was independent of the jet direction;

\item the transverse momentum of the 
jets, $P^*_{T,\rm jet}$,  measured with respect to the $\gamma^* \Pma$ axis, was reconstructed 
without  significant bias and with a resolution of 0.7 GeV;

\item the jet pseudorapidity, $\eta^{\rm jet}_{\rm lab}$,  measured
 in the laboratory frame (see below), was reconstructed without 
 significant bias and with a resolution of 0.1 to 0.2 units.

\end{itemize}

In addition, it was checked that, in the $M_X$ region considered and for a 
fixed value of $y_{\rm cut} = 0.05$, the jets, defined in the centre-of-mass frame,  
always had energy significantly above the noise level in the CAL.
The jet energies in the laboratory frame, $E^{\rm jet}_{\rm lab}$, 
 computed by boosting the jet four-momenta in 
the centre-of-mass system back  to the  laboratory frame, were 
typically above 4~GeV, thus ensuring  a well understood 
jet-energy reconstruction for all jets. \\ 

The sample of three-jet final states was  defined by using a fixed $y_{\rm cut} = 0.05$ and the 
jets were required to be in the range  $|\eta^{\rm jet}_{\rm lab}| < 2.3$. 
The variable  $\eta^{\rm jet}_{\rm lab}$ was
computed in the same way as $E^{\rm jet}_{\rm lab}$.   
The effect of the inclusion of jet masses in the 
determination of $\eta^{\rm jet}_{\rm lab}$ was found to be negligible. 
This procedure selected events with jets reconstructed in the central part of the 
detector, which are not affected by the   
$\eta^{\rm max}$ cut applied in the forward region and for which a 
good correlation between the jet before and after
detector effects was  preserved.
After the $\eta^{\rm jet}_{\rm lab}$ cut, a final sample of 891 events with exactly   
three identified jets was obtained.


\subsection{Three-jet  topology variables}
Neglecting jet mass, the topology of a three-jet final 
state can be described using the fractional-momentum variables 
\begin{equation}
x_i = \frac{2 \cdot |{\bf{P}}^{\rm jet}_i|}{\sum_j{|{\bf{P}}^{\rm jet}_j|}}, \ \ \ \ \ \ |{\bf{P}}_i^{\rm jet}| = \sqrt{(P^{\rm jet}_{i,X})^2+(P^{\rm jet}_{i,Y})^2+(P^{\rm jet}_{i,Z})^2}\  ,
\label{eq:xi}
\end{equation}
where ${\bf{P}}^{\rm jet}_i \ \  (i=1-3)$ denotes the  $i^{\rm th}$ jet 
three-momentum, determined in the centre-of-mass frame, 
and by construction $x_1+x_2+x_3=~2$. The jets were sorted according to their 
momenta  in such a way that
$$
x_1 \geq \ x_2 \ \geq x_3 \  . 
$$
Two of the $x_i$ variables in Eq.~(\ref{eq:xi}) are sufficient to describe the jet topology in 
 the final state. Following the studies of three-jet production in $e^+ e^-$ collisions
 at PETRA \cite{jade,tasso}, these quantities were chosen to be 
$x_1$ and $\xi = (x_2 - x_3)/x_1$, as suggested by Ellis and Karliner \cite{x1x2x3}. 

\subsection{Jet-shape reconstruction}

The internal structure of the jets was studied in terms 
of the differential jet shape, defined in the centre-of-mass system 
as the average of the fraction of the jet energy which lies inside 
an annulus of inner angular distance $\varphi -\delta \varphi/2$  and 
outer angular distance $\varphi+\delta \varphi/2$  from 
the jet axis. EFOs were used to reconstruct the jet shape defined as
$$
\rho^{\rm EFOs} (\varphi) = \frac{1}{\delta \varphi \ N_{\rm jets}} \sum_{\rm jets} \frac{\Delta E^*_{\rm jet}(\varphi \pm \delta \varphi/2)}{E^*_{\rm jet}} \ ,
$$
where $\Delta E^*_{\rm jet}(\varphi \pm \delta \varphi/2)$  denotes the sum of the energies of the EFOs 
belonging to a given jet whose angular distance to the jet axis is within the 
range $(\varphi-\delta \varphi/2,\varphi+ \delta \varphi/2)$ with 
$\delta \varphi= 0.2$ radians; $N_{\rm jets}$ is the number of jets. For MC events, the same 
jet-shape definition as used above for the EFOs was applied to the final-state
hadrons.


\section{Unfolding and systematic studies}

The measured distributions were corrected bin-by-bin for acceptance and 
migrations using RAPGAP-CDM. The measurement was carried out in the region 
$5 < Q^2 < 100$ GeV${}^2$, $200 < W < 250$ GeV,
 $23 < M_X < 40$ GeV and $x_{\Pma}<0.025$, for diffractive events with 
$\eta^{\rm max}_{\rm hadron} < 3.0$ and $|\eta^{\rm jet}_{\rm lab}|_{\rm hadron} < 2.3$, 
where $\eta^{\rm max}_{\rm hadron}$ is the pseudorapidity of the most-forward final-state 
hadron with energy above 400 MeV.
The measurements presented here thus cover a very restricted phase-space region.
According to RAPGAP, in the kinematic region considered, less than $10 \%$ of 
all diffractive events have  $\eta^{\rm max}_{\rm hadron} < 3.0$ and three 
jets in the final state with $|\eta^{\rm jet}_{\rm lab}|_{\rm hadron} < 2.3$.
The fraction of three-jet events  selected by the  $\eta^{\rm max}_{\rm hadron} < 3.0$ and 
$|\eta^{\rm jet}_{\rm lab}|_{\rm hadron} < 2.3$ requirements was estimated using RAPGAP to 
be of the order of $30 \%$.\\

A detailed study of the main sources contributing to the systematic uncertainties of the measurements 
was performed \cite{taro}. The different contributions are listed below, with a typical value quoted for each:
\begin{itemize} 

\item  the uncertainty in the hadronic final-state energy scale 
was taken into account by varying 
the CAL (FPC) energy by $\pm 3 \%$ ($\pm 10 \%$) 
in the MC, after removing the energy deposits belonging  to the 
scattered positron. This introduced an uncertainty in the measured distributions 
of about  $5\%$; 

\item the selected window in the reconstructed $M_X$  was
varied by $\pm 10 \%$, both in data and simulated events, according to the estimated 
resolution in the reconstruction of this quantity.  The measured cross sections 
varied by $5\%$;    

\item the uncertainty in the positron energy scale 
was included by varying the positron energy by $\pm 2 \%$ in the MC.
The effect on the $\eta^*_{\rm jet}$ and $P^*_{T, \rm jet}$ distributions (see Section 8)
was typically below $3\% $;

\item  the box cut applied to the position of the scattered positron in the RCAL  
was varied by $\pm 5$ mm (both in data and simulated events) 
to account for the uncertainty in simulating the measured positron position. The effect was 
typically below $1 \%$;

\item the selected range in the reconstructed jet pseudorapidity in the laboratory frame,   
$|\eta^{\rm jet}_{\rm lab}|<2.3$, (see Section 6.1) was varied by 0.2 units
around the nominal value,  both in data and MC simulation, 
according to the estimated resolution in the reconstruction of the 
jet direction. This introduced an uncertainty on the measured cross sections 
of about $2\%$;

\item the measured distributions were corrected to the hadron level using RAPGAP-PS  instead of
RAPGAP-CDM  to account for the uncertainty in the simulation of higher-order QCD
processes in the MC model. The effect varied between $5\%$ and $10 \%$ and was the dominant systematic uncertainty.

\end{itemize}

The final systematic uncertainty was computed by adding 
the different contributions in quadrature. The contribution from proton-dissociative processes,
 $e p \rightarrow e X N$, with $M_N \leq 3$~GeV, where the system $N$ escaped through the 
forward beam hole in the FPC, was estimated using EPSOFT to be $(16 \pm 5)\%$ and was subtracted. 
The overall normalisation uncertainty of $5.4 \%$ in the measured cross sections, coming  
from the uncertainty on the proton-dissociative contribution and  the $2 \%$ error in the luminosity 
determination, is not included in the figures.

\section{Results}
 Multi-jet production in the centre-of-mass system of the hadronic
 final state, $X$, is presented in the kinematic region defined in Section 7 
 using the exclusive $k_T$-algorithm with $y_{\rm cut} = 0.05$.
The measurements\footnote{The measurements are tabulated in Tables 1 - 7.} 
were corrected for electroweak radiative effects to the Born level.   

\subsection{Three-jet  topology}

\begin{figure}
\epsfysize=21cm
\epsffile{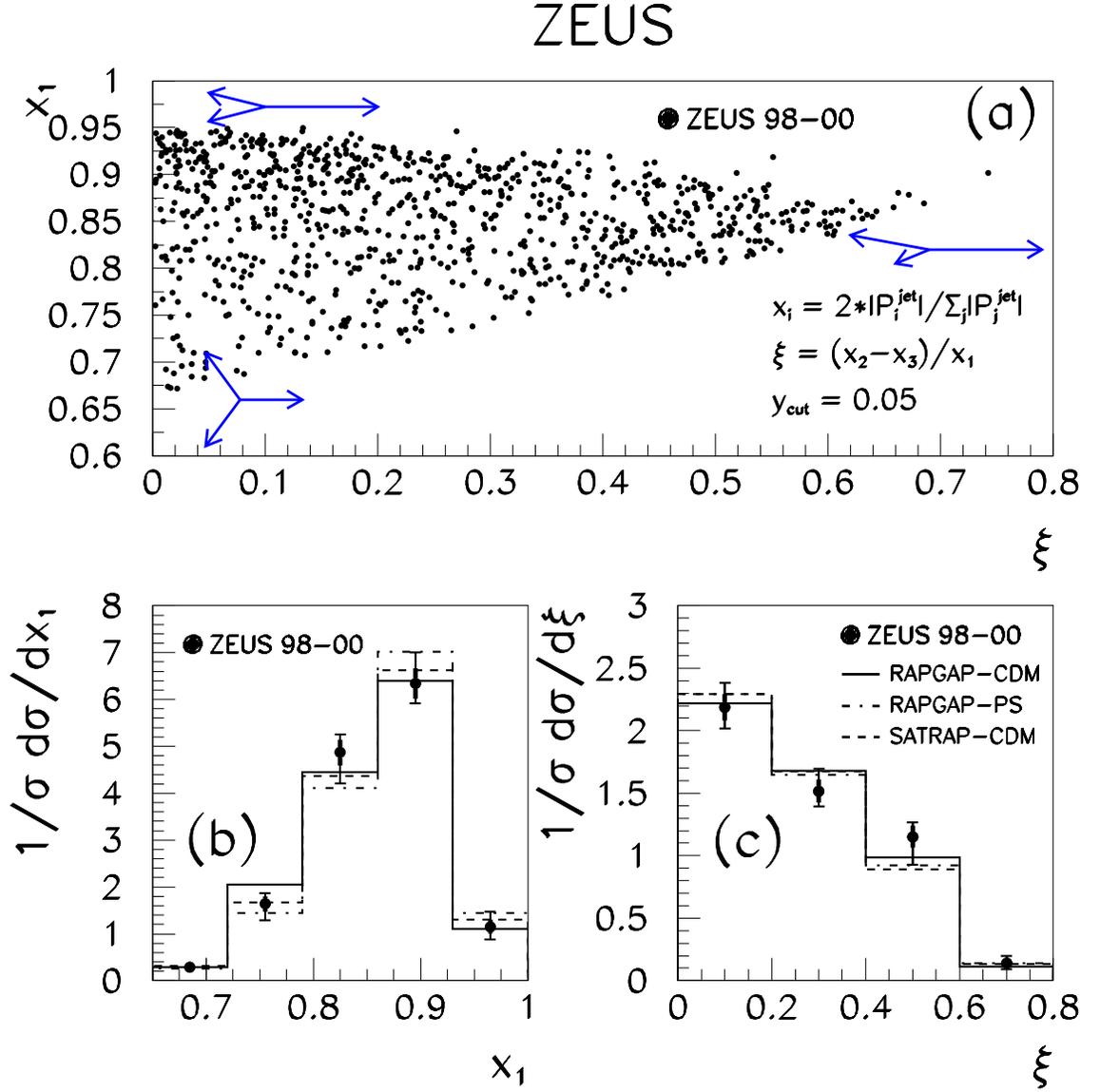}
\vspace{-2cm}
\caption{\label{fig2}{
(a) Distribution of the  three-jet sample 
in the $(\xi, \ x_1)$ plane. The different topologies within 
the plane are indicated, where the length of the arrow represents the 
momentum of the jet as computed in the centre-of-mass frame. (b,c)  
Normalised differential cross sections as a function of $x_1$ and  
$\xi$ in three-jet production. The thick error bars indicate 
the statistical uncertainties and the 
thin error bars indicate the statistical and systematic uncertainties 
combined in quadrature. The MC expectations 
from {RAPGAP-CDM} (solid line),  RAPGAP-PS (dashed-dotted line) and  
SATRAP-CDM  (dashed line) are shown.
}
}
\end{figure}

The topology of the three-jet final state was studied using $x_i$, the fractional-momentum variables (see Section 6.2). 
Figure \ref{fig2}(a) shows the distribution of the three-jet 
sample in the ($\xi  , \  x_1$) plane. As  indicated in the figure, different 
regions in the plane correspond to different three-jet topologies. All   
configurations are present, including those for which the three jets have similar energies.     
The distribution of the events  shows that, for the given $y_{\rm cut}$, 
  the favoured  three-jet topologies are those for which 
one leading jet carries approximately twice the energy of either of the other two 
jets $(\xi \simeq 0.1 , x_1 \simeq 0.9)$. The normalised differential cross 
sections for three-jet production as a function of $x_1$ and  $\xi$ 
are presented in Figs. \ref{fig2}(b) and \ref{fig2}(c), compared to the MC predictions. 
The measured $x_1$ distribution has a peak structure with a maximum at $x_1 \simeq 0.9$,  
which is determined by the $y_{\rm cut}$ value used in defining jets since the position of the peak  
approximately corresponds to $1 - 2 y_{\rm cut}$.  
The shapes of the measured cross sections are 
reproduced by all three MC models. This indicates that they provide a
good description of the dynamics of the three-jet emission, which dictates the population of 
the ($\xi , \  x_1$) plane.\\

\begin{figure}
\epsfysize=21cm
\hspace{-1.5 cm}
\epsffile{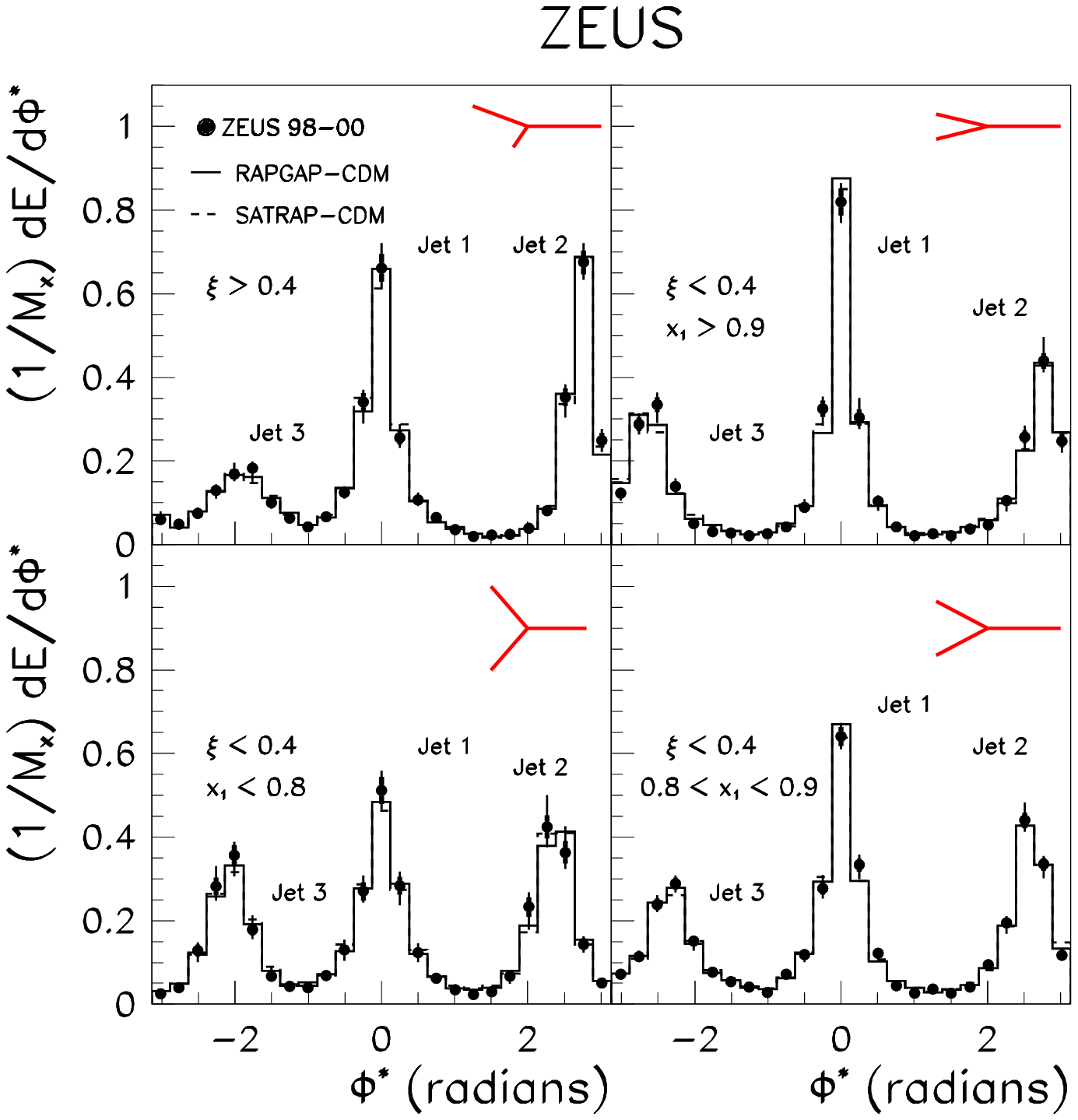}
 \vspace{-3cm}
\caption{\label{fig3}{ 
Energy-flow distribution in the three-jet plane, normalised to the total invariant mass,  
as a function of the azimuthal angle, $\phi^*$,  in the different regions 
of the ($\xi ,x_1$) plane. The azimuthal angle is defined
 to run from the first to the second most-energetic jet in each event. 
 The thick error bars indicate the statistical uncertainties and the 
thin error bars indicate the statistical and systematic uncertainties 
combined in quadrature. The MC expectations 
from RAPGAP-CDM (solid lines) and SATRAP-CDM  (dashed lines) are shown.
The characteristic topology of the three-jet configuration is indicated in the 
top right corner of each plot.
}
}
\end{figure}

The energy  flow, as measured in the event plane
defined by the two most-energetic jets in the event, was studied as a function
of the azimuthal angle, $\phi^*$. 
The origin of $\phi^*$ was defined to be along the direction of the most-energetic jet with
positive $\phi^*$ increasing in the direction 
of the second most-energetic jet.  The energy  flow 
was measured by projecting the particle momenta onto the event plane and 
summing, in each $\phi^*$ bin,  the particle energies.
The sums were normalised by dividing by $M_X$.
In Fig.~\ref{fig3},  the normalised energy flow 
is presented for different regions of the   ($\xi , \ x_1$) plane. 
A clear three-jet structure is observed in all regions. 
 The measurements are compared with the RAPGAP and SATRAP-CDM   
 predictions.  RAPGAP-CDM, RAPGAP-PS (not shown but similar to  RAPGAP-CDM) 
 and SATRAP-CDM all provide a very good description of the
 measured flows for all ($\xi , \ x_1$) regions and $\phi^*$ ranges. 

\subsection{Cross section as a function of {\boldmath{$\eta^*_{\rm jet}$}}}

Figure~\ref{fig4}(a,b) shows the differential cross section as a function 
of the jet pseudorapidity, ${d\sigma}/{d\eta^*_{\rm jet}}$, measured with respect to  
the $\gamma^* \Pma$ axis. All jets in each event were included. The total cross section for three-jet 
production, obtained by integrating the differential measurement, is:
$$
\sigma^{\rm 3 jets} = \frac{1}{3} \cdot \int \frac{d\sigma}{d\eta^*_{\rm jet}} \ {d\eta^*_{\rm jet}} = 14.1 \pm 0.5 (\rm stat.) {}^{+ 1.3}_{- 1.1}(\rm syst.) \  \rm pb.
$$
For this measurement, the overall normalisation uncertainty of $5.4 \%$ (see Section 7)  
is combined in quadrature in the quoted total systematic uncertainty.
The measured cross section corresponds to $32.2\% \pm 0.9 \% (\rm stat.) {}^{+ 1.3}_{-1.6} \% (\rm syst.)$ of the
diffractive cross section for two or more jets. This rate  
is consistent with being independent of $Q^2$ and $M_X$.\\

Figure~\ref{fig4}(a) shows the measured differential cross 
section compared to the RAPGAP-CDM and RAPGAP-PS predictions. Both implementations of        
RAPGAP provide a good description of the shape of the cross section but  
underestimate it by about $20 \%$. 
The observed deficit in the absolute normalisation predicted by RAPGAP 
may be attributed to the absence of next-to-leading-order 
terms in the computed matrix elements, which are only approximated via parton 
cascades.\\

In Fig.~\ref{fig4}(b) the measured differential cross section is 
compared to the SATRAP, SATRAP-CDM and RIDI predictions. 
For the latter, the contribution from the hadron jet  
associated with the gluon from the  $q\overline{q}g$ final state is shown separately.
The {SATRAP} and {RIDI} predictions are 
found to be more than a factor of two below the measured cross sections and    
 show a double-peak structure not  observed in the data. This tendency is 
particularly pronounced in RIDI. This reflects the presence of 
dominant topologies in these models in which 
the jet in the Pomeron direction is emitted almost collinear to the  
$\gamma^* \Pma$ axis and for which the second and third jets are found 
in the photon hemisphere. The observed shape of the {SATRAP} and {RIDI} predictions 
is a consequence of the inadequacies of the implementation of the modelling of 
higher-order QCD processes in these MC models as discussed in Section 4. 
SATRAP-CDM describes the shape of the measured distribution
but also underestimates the normalisation by about $20 \%$, which, as for RAPGAP, may  
be   attributed to the absence of  next-to-leading-order and some leading-order 
terms in the computed matrix elements. 
   
\subsection{{\boldmath{$P^*_{T, \rm  jet}$}} distribution for the most-forward jet}
 
\begin{figure}
\epsfysize=21cm
\epsffile{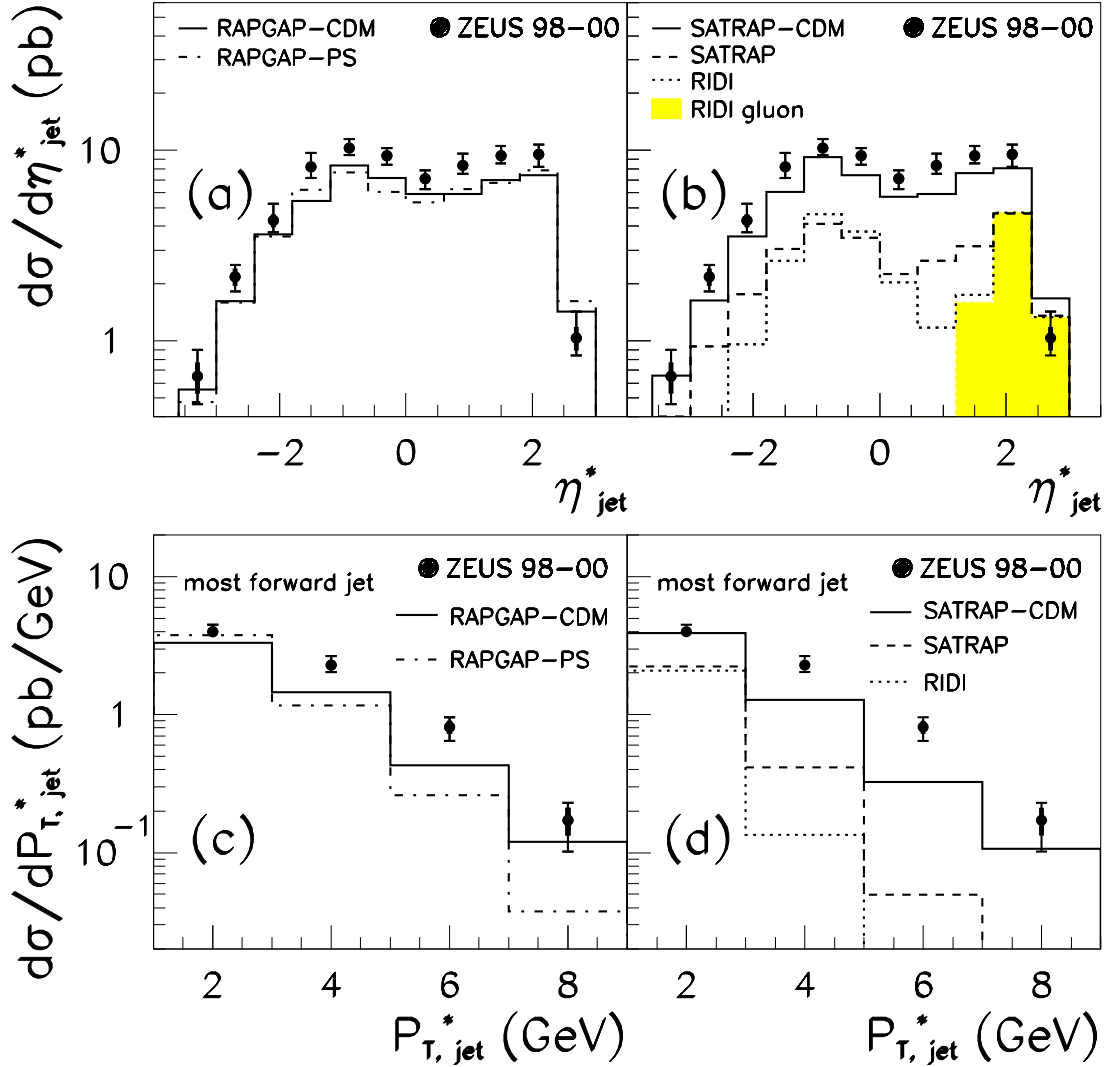}
\vspace{-3cm}
\caption{\label{fig4}{ 
Differential cross section for three-jet production
as a function of (a,b) the jet pseudorapidity,  
$d\sigma/d\eta^*_{\rm jet}$, of the 
three jets; (c,d) the jet transverse 
momentum, $P^*_{\rm T, jet}$, of the most-forward jet.
The Pomeron direction defines the forward hemisphere. 
The thick error bars indicate the statistical uncertainties and the 
thin error bars indicate the statistical and systematic uncertainties 
combined in quadrature.
The overall normalisation uncertainty of $5.4 \%$ was not 
included in the figures.
The MC expectations are shown 
from (a,c) {RAPGAP-CDM} (solid line) and RAPGAP-PS (dashed-dotted line) and (b,d) SATRAP  
(dashed line), {RIDI} (dotted line) and SATRAP-CDM (solid line). For 
RIDI, the contribution from the gluon-initiated jet is 
shown separately in (b) (shaded area). 
}
}
\end{figure}

The differential cross section as a function of the 
jet transverse momentum, $P^*_{T,\rm jet}$, measured with respect to the $\gamma^* \Pma$ axis  
for the most-forward jet is presented in Fig.~\ref{fig4}(c,d). 
The measurement was performed  in the region $P^*_{T,\rm jet} > 1$ GeV, where 
 a good parton jet-hadron jet correlation is preserved, and for 
which the measurement is 
well defined, given the   
estimated resolution in the reconstruction of $P^*_{T,\rm jet}$ in the detector (see Section 6).
The measured cross section is dominated by configurations 
with $P^*_{T,\rm jet}$  below 3 GeV but contains a long tail towards 
larger $P^*_{T,\rm jet}$ values, reflecting the presence of contributions from
higher-order QCD processes.\\

Figure~\ref{fig4}(c) shows the  measured cross section compared to RAPGAP-CDM and RAPGAP-PS. 
The RAPGAP predictions underestimate the measured cross section  
and have  a slightly different shape than the data. RAPGAP-CDM 
provides a better description of the data than RAPGAP-PS. 
The observed difference
between these two predictions indicates the sensitivity of the 
$P^*_{T,\rm jet}$ distribution to the approach used for the 
simulation of higher-order QCD processes in the MC.\\

 Figure~\ref{fig4}(d) shows the measured $P^*_{T,\rm jet}$ distribution  
compared to the predictions from  SATRAP,  SATRAP-CDM and RIDI.  The $P^*_{T,\rm jet}$ distribution 
predicted by SATRAP and RIDI  is steeper than that of the data, indicating that,  
 in these MC models, the large $P^*_{T,\rm jet}$ configurations are missing. The cross section 
in these models is dominated by configurations strongly aligned with respect 
to the $\gamma^* \Pma$ axis in the forward region. As already mentioned, 
this is most probably a consequence of the inadequacies of the implementation of the modelling of 
higher-order QCD processes in these models. In SATRAP-CDM, the shape and 
normalisation of the $P^*_{T,\rm jet}$ distribution is in better  agreement with the data.\\

In all models, the gluon in the 
$q\overline{q}g$ final state is emitted 
with very small transverse momentum with respect to the $\gamma^* \Pma$ axis. 
The contributions from higher-order QCD processes  
are expected to be important and to influence the kinematics of the corresponding hadron 
jet which, due to string effects, tends to have smaller pseudorapidity (larger transverse momentum) compared to 
the initial parton. Given the $|\eta^{\rm jet}_{\rm lab}|_{\rm hadron}< 2.3$ requirement applied in defining the 
measured cross sections, the  absolute predictions  from these models depend  on 
an appropriate treatment of these higher-order QCD emissions and a proper 
description of the forward hadron-jet dynamics (as shown in Fig.~\ref{fig4}(d)).
This is particularly clear in the case of SATRAP, which  fails to reproduce the three-jet data but 
provides a good description of inclusive diffractive DIS~\cite{satrap}.

\subsection{Jet shapes}
\begin{figure}
\epsfysize=21cm
\epsffile{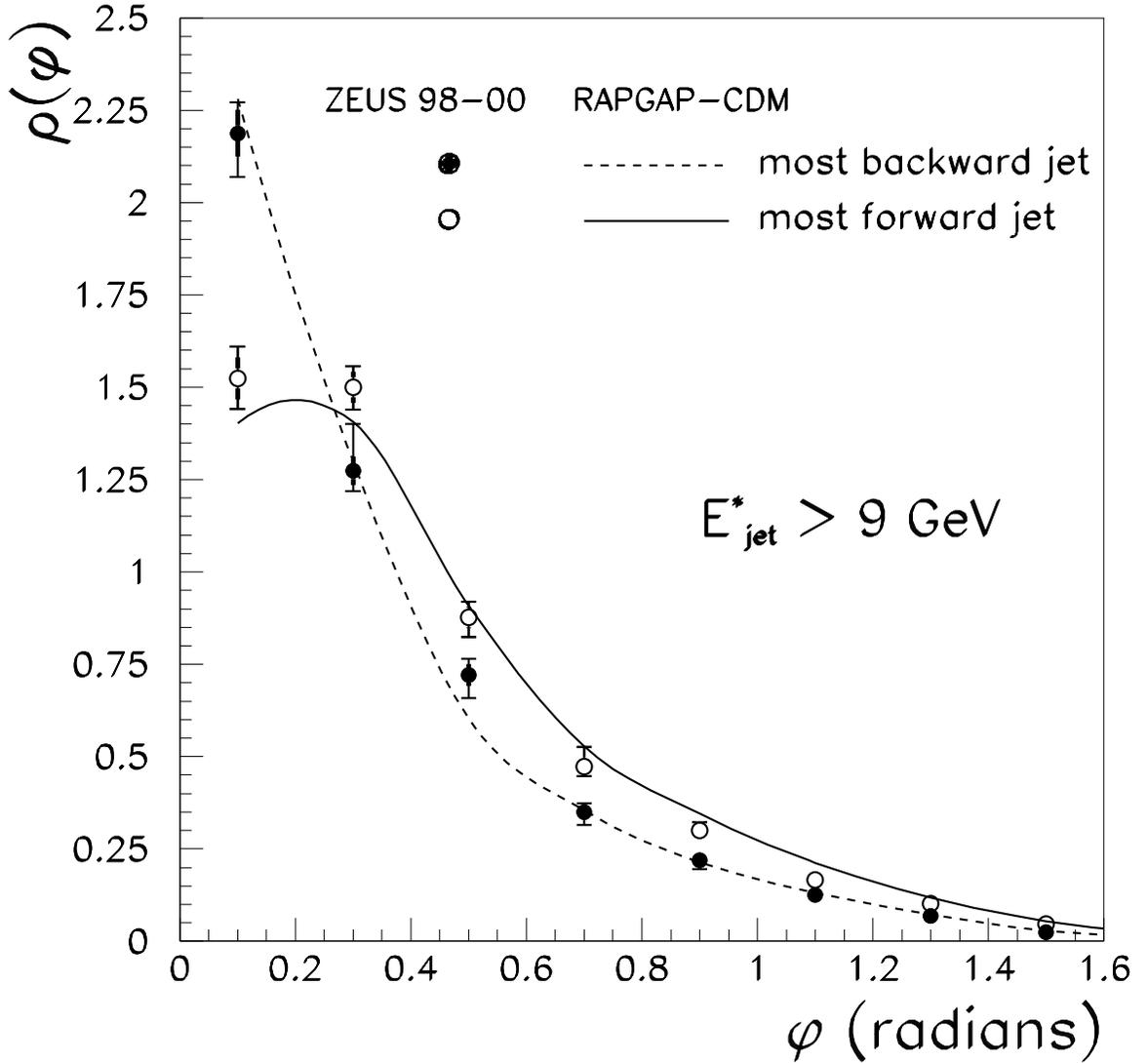}
\vspace{-2cm}
\caption{\label{fig5}{ 
The differential jet shape, $\rho(\varphi)$, for the most-forward and 
most-backward jets  with $E^*_{\rm jet} >$ 9 GeV in three-jet events, where the 
Pomeron defines the forward direction.  
The thick error bars indicate the statistical uncertainties and the 
thin error bars indicate the statistical and systematic uncertainties 
combined in quadrature.
The MC expectations 
from RAPGAP-CDM are shown. 
}
}
\end{figure}

The differential jet shapes, $\rho(\varphi)$, were measured 
for the most-forward and most-backward jet in three-jet events, where the forward region 
is defined by the Pomeron direction. The measurements were performed in the $\gamma^* \Pma$ centre-of-mass frame for 
jets with energy, $E^*_{\rm jet}$, above 9~GeV. 
This requirement removes
the contributions from the least-energetic  jet in the event which, 
in the $M_X$-range considered and for the given $y_{\rm cut}$, has energy 
below this threshold and for which hadronisation effects on the jet shape are 
expected to be important.   
Figure \ref{fig5} shows the measured jet shapes compared with the 
RAPGAP-CDM predictions (the SATRAP-CDM predictions, not shown, are almost identical). The jet 
in the Pomeron direction is broader than the jet in the photon direction.
These measurements are  described by  models 
in which a gluon populates the Pomeron hemisphere and a quark
is found in the photon direction. The energy distribution of the two jets 
is similar, indicating that the 
observed difference in the measured jet shapes can be ascribed  to the different nature of the 
initial partons. Moreover, the 
difference between the jet shapes in the Pomeron and photon 
directions is qualitatively similar to  that between quark- and gluon-initiated
jets as measured in $e^+ e^-$ experiments \cite{opalshape}. This measurement 
therefore supports  the validity of the picture where the three-body final state 
is dominated by a $q\overline{q}g$  configuration with the gluon  
preferentially emitted in the Pomeron direction.

\section{Summary and conclusions}
 Multi-jet production 
has been studied in the kinematic region 
 $5 < Q^2 < 100$ GeV${}^2$, $200 < W < 250$ GeV,
 $23 < M_X < 40$ GeV and $x_{\Pma}<0.025$, for diffractive events with 
$\eta^{\rm max}_{\rm hadron} < 3.0$ and $|\eta^{\rm jet}_{\rm lab}|_{\rm hadron} < 2.3$.
A study of three-jet production, using the exclusive $k_T$-algorithm with $y_{\rm cut} = 0.05$
in the centre-of-mass system of the hadronic final state,  
has been presented. The rate of three-jet production has been 
measured to be $32.2 \% \pm 0.9 \% (\rm stat.) {}^{+ 1.3}_{-1.6} \% (\rm syst.)$ of the diffractive 
jet cross section, and is consistent with being independent of $Q^2$ and $M_X$. 
The topologies of the three-jet final states 
have been studied in terms of the jet fractional-momentum 
variables $\xi $ and $x_1$; all possible 
topologies have been observed.  
The distribution of the events in the ($\xi  , \  x_1$)  plane shows that, for a given $y_{\rm cut}$, 
  the favoured topologies are those for which one leading jet carries 
approximately twice the energy of either of the other two jets $(\xi \simeq 0.1 , x_1 \simeq 0.9)$.
Differential cross sections for three-jet 
production have been measured as a function of the jet pseudorapidity, $\eta^*_{\rm jet}$, 
 and jet transverse momentum, $P^*_{T,\rm jet}$, with respect to the $\gamma^* \Pma$ axis.
The total cross section for three-jet production, determined by integrating the 
$\eta^*_{\rm jet}$ distribution, is  
$\sigma^{\rm 3 jets} = 14.1 \pm 0.5 (\rm stat.) {}^{+ 1.3}_{-1.1}(\rm syst.) \ \rm pb$.
The $P^*_{T,\rm jet}$ distribution for the most-forward jet  
is dominated by configurations with $P^*_{T,\rm jet}$ below 3 GeV but exhibits  
 a  long tail towards larger values, reflecting the presence of 
higher-order QCD contributions.   
The differential jet shapes, $\rho (\varphi)$, of the two most-energetic jets in three-jet events  
have been measured. The jet in the 
Pomeron direction is broader and  consistent with  the shape of   
gluon-initiated jets as predicted by diffractive MC models.\\


The data are broadly consistent with models in which the hadronic 
final-state is dominated  
by a $q\overline{q}g$ system with the gluon preferentially emitted in the Pomeron direction.  
Such configurations are predicted both by resolved Pomeron models (as implemented in RAPGAP)
 with a Pomeron dominated by  gluons, and by models where the virtual photon 
dissociates in a $q\overline{q}g$ system which interacts with the proton via 
the exchange of a gluon ladder (as implemented in SATRAP and RIDI). 
RAPGAP provides a reasonable description of the measured distributions, although it underestimates the
 measured cross section by about $20 \%$. This deficit may be attributed to the 
absence of next-to-leading-order corrections, which are only included approximately via 
parton cascades.
The predictions from SATRAP and RIDI fail to describe the shape and normalisation of the 
$\eta^*_{\rm jet}$ and  $P^*_{T,\rm jet}$ distributions.
However, it has been shown that an improved implementation of the modelling of higher-order 
QCD processes in SATRAP-CDM gives reasonable agreement with the data. 
Thus, the study of three jets in diffraction constitutes a very 
sensitive method to investigate the dynamics of diffractive deep inelastic scattering.

\vspace{0.5cm}
\noindent {\Large\bf Acknowledgements}
\vspace{0.3cm}

 The strong support and encouragement of the DESY Directorate have been 
invaluable. The experiment was made possible by the inventiveness and the 
diligent efforts of the HERA machine group.  The design, construction and 
installation of the ZEUS detector have been made possible by the
ingenuity and dedicated efforts of many people from inside DESY and
from the home institutes who are not listed as authors. Their 
contributions are acknowledged with great appreciation. 
We would like to thank the staff support from the various 
Institutes which collaborated in the construction of the FPC and
in the setup of the test systems, in particular
J. Hauschildt and  K. L\"offler (DESY), L. Herv\'as (CERN), 
R. Feller, E. M\"oller and H. Prause (I. Inst. for Exp. Phys., Hamburg),
A. Maniatis (II. Inst. for Exp. Phys., Hamburg), 
and the members of the mechanical workshop of the Faculty of Physics 
from Freiburg University.


\clearpage


\clearpage

\renewcommand{\baselinestretch}{1.12}
\renewcommand{\arraystretch}{1.1}
\parindent       0.0pt
\parskip         0.3cm plus0.05cm minus0.05cm
\overfullrule    0.0pt
\itemsep    0.2ex plus 0.1ex minus 0.15ex
\topsep     0.2ex plus 0.1ex minus 0.15ex
\partopsep  1.2ex plus 0.6ex minus 0.6ex
\setlength\leftmargini   {1.6em}
\setlength\leftmarginii  {1.4em}
\setlength\leftmarginiii {1.2em}
\setlength\leftmarginiv  {1.0em}


\small


\begin{table}
\begin{center}
\begin{tabular} {||c||c|c||c|c||c|c||}      \hline
\multicolumn{7}{|c|}{$N_{\rm jet}$ fractions versus $y_{\rm cut}$} \\ \hline\hline
\multicolumn{1}{|c||}{} & \multicolumn{2}{c||}{$N_{\rm jet} = 2$} &\multicolumn{2}{c||}{$N_{\rm jet} = 3$} & \multicolumn{2}{c||}{$N_{\rm jet} > 3$} \\ \hline\hline
$y_{\rm cut}$ & rate & (stat. $\bigoplus$ syst.) & rate & (stat. $\bigoplus$ syst.) &rate & (stat. $\bigoplus$ syst.) \\ \hline
0.01& $0.232$&${}_{-0.010}^{+0.017}$ & $ 0.471$&${}_{-0.012}^{+0.008} $ & $ 0.297$&${}_{-0.009}^{+0.007} $ \\ \hline
0.02& $0.503$&${}_{-0.016}^{+0.023}$ & $ 0.420$&${}_{-0.024}^{+0.016} $ & $ 0.077$&${}_{-0.004}^{+0.004} $ \\ \hline
0.03& $0.655$&${}_{-0.012}^{+0.021}$ & $ 0.320$&${}_{-0.022}^{+0.011} $ & $ 0.025$&${}_{-0.002}^{+0.002} $ \\ \hline
0.04& $0.750$&${}_{-0.010}^{+0.012}$ & $ 0.241$&${}_{-0.012}^{+0.010} $ & $ 0.009$&${}_{-0.001}^{+0.001} $ \\ \hline
0.05& $0.809$&${}_{-0.009}^{+0.011}$ & $ 0.187$&${}_{-0.011}^{+0.009} $ & $ 0.004$&${}_{-0.001}^{+0.001} $ \\ \hline
0.06& $0.850$&${}_{-0.007}^{+0.011}$ & $ 0.149$&${}_{-0.011}^{+0.007} $ & $ 0.001$&${}_{-0.000}^{+0.001} $ \\ \hline
0.07& $0.878$&${}_{-0.006}^{+0.009}$ & $ 0.122$&${}_{-0.008}^{+0.006} $ &  -  &  -  \\ \hline
0.08& $0.902$&${}_{-0.006}^{+0.007}$ & $ 0.097$&${}_{-0.007}^{+0.005} $ &  -  &  -  \\ \hline
0.09& $0.919$&${}_{-0.004}^{+0.006}$ & $ 0.081$&${}_{-0.006}^{+0.004} $ &  -  &  -  \\ \hline
0.10& $0.934$&${}_{-0.004}^{+0.004}$ & $ 0.066$&${}_{-0.004}^{+0.004} $ &  -  &  -  \\ \hline
\end{tabular}
\caption{Measurement of the $n_{\rm jet}$-fraction, for $n_{\rm jet}=2,3$ and $ > 3$, 
as a function of the exclusive $k_T$ jet resolution parameter, $y_{\rm cut}$.}
\label{table1}
\end{center}
\end{table}


\begin{table}
\begin{center}
\begin{tabular} {||c||c|c|c||}      \hline
\multicolumn{4}{|c|}{{\Large{$\frac{1}{\sigma} \frac{d\sigma}{d x_1}$}}}\\ \hline\hline
$d x_1$ & $\frac{1}{\sigma} \frac{d\sigma}{d x_1}$ & (stat.) & (stat. $\bigoplus$ syst.) \\ \hline\hline
 [0.650,0.720]&    0.288&   $\pm    0.056$&   ${}_{-0.091}^{+ 0.098}$ \\ \hline
 [0.720,0.790]&    1.639&   $\pm    0.162$&   ${}_{-0.345}^{+ 0.226}$ \\ \hline
 [0.790,0.860]&    4.869&   $\pm    0.277$&   ${}_{-0.656}^{+ 0.387}$ \\ \hline
 [0.860,0.930]&    6.339&   $\pm    0.319$&   ${}_{-0.428}^{+ 0.668}$ \\ \hline
 [0.930,1.000]&    1.150&   $\pm    0.150$&   ${}_{-0.266}^{+ 0.319}$ \\ \hline
\end{tabular}
\caption{Normalised differential cross section as a function of $x_1$.}
\label{table2}
\end{center}
\end{table}

\begin{table}[p]
\begin{center}
\begin{tabular} {||c||c|c|c||}      \hline
\multicolumn{4}{|c|}{{\Large{$\frac{1}{\sigma} \frac{d\sigma}{d \xi}$}}}\\ \hline\hline
$d \xi$ & $\frac{1}{\sigma} \frac{d\sigma}{d \xi}$ & (stat.) & (stat. $\bigoplus$ syst.) \\ \hline\hline
 [0.0,0.2] &   2.189&   $\pm    0.106$&  ${}_{-0.169}^{+0.196}$ \\ \hline 
 [0.2,0.4] &   1.516&   $\pm    0.093$&  ${}_{-0.123}^{+0.179}$ \\ \hline
 [0.4,0.6] &   1.153&   $\pm    0.086$&  ${}_{-0.225}^{+0.114}$ \\ \hline
 [0.6,0.8] &   0.142&   $\pm    0.033$&  ${}_{-0.051}^{+0.057}$ \\ \hline
\end{tabular}
\caption{Normalised differential cross section as a function of $\xi$.}
\label{table3}
\end{center}
\end{table}


\begin{table}
\begin{center}
\begin{tabular} {||c||c|c||c|c||c|c||c|c||}      \hline
\multicolumn{9}{|c|}{{\Large{$\frac{1}{M_X} \frac{dE}{d\phi^*}$}}}\\ \hline\hline
 \multicolumn{1}{|c}{}& \multicolumn{2}{|c|}{$ \xi > 0.4 $} &  \multicolumn{2}{|c|}{$\xi < 0.4$, $\ x_1 > 0.9$} &  \multicolumn{2}{|c|}{$ \xi < 0.4$, $\ x_1 < 0.8$} &  \multicolumn{2}{|c|}{$\xi < 0.4$, $0.8 < x_1 < 0.9$} \\ \hline\hline
$\overline{\phi^*}$ (rad.) & $\frac{1}{M_X} \frac{dE}{d\phi^*}$ & {\tiny{(stat. $\bigoplus$ syst.)}} & $\frac{1}{M_X} \frac{dE}{d\phi^*}$ & {\tiny{(stat. $\bigoplus$ syst.)}} & $\frac{1}{M_X} \frac{dE}{d\phi^*}$ & {\tiny{(stat. $\bigoplus$ syst.)}} & $\frac{1}{M_X} \frac{dE}{d\phi^*}$ & {\tiny{(stat. $\bigoplus$ syst.)}} \\ \hline

  -3.016 & $ 0.061$&${}_{-0.008}^{+0.019} $ & $0.122$&${}_{-0.017}^{+0.014} $ & $0.024$&${}_{-0.005}^{+0.006} $ & $0.072$&${}_{-0.012}^{+0.008}$ \\ \hline
  -2.765 & $ 0.049$&${}_{-0.011}^{+0.009} $ & $0.288$&${}_{-0.024}^{+0.026} $ & $0.040$&${}_{-0.010}^{+0.007} $ & $0.114$&${}_{-0.011}^{+0.015}$ \\ \hline
  -2.513 & $ 0.075$&${}_{-0.012}^{+0.013} $ & $0.335$&${}_{-0.043}^{+0.029} $ & $0.129$&${}_{-0.028}^{+0.019} $ & $0.239$&${}_{-0.018}^{+0.023}$ \\ \hline
  -2.262 & $ 0.130$&${}_{-0.020}^{+0.014} $ & $0.139$&${}_{-0.014}^{+0.020} $ & $0.282$&${}_{-0.033}^{+0.048} $ & $0.289$&${}_{-0.022}^{+0.020}$ \\ \hline
  -2.011 & $ 0.168$&${}_{-0.015}^{+0.028} $ & $0.050$&${}_{-0.007}^{+0.010} $ & $0.356$&${}_{-0.049}^{+0.034} $ & $0.151$&${}_{-0.023}^{+0.012}$ \\ \hline
  -1.759 & $ 0.182$&${}_{-0.036}^{+0.016} $ & $0.031$&${}_{-0.006}^{+0.005} $ & $0.179$&${}_{-0.023}^{+0.033} $ & $0.077$&${}_{-0.008}^{+0.010}$ \\ \hline
  -1.508 & $ 0.100$&${}_{-0.014}^{+0.016} $ & $0.027$&${}_{-0.006}^{+0.007} $ & $0.068$&${}_{-0.013}^{+0.020} $ & $0.054$&${}_{-0.009}^{+0.008}$ \\ \hline
  -1.257 & $ 0.063$&${}_{-0.012}^{+0.013} $ & $0.021$&${}_{-0.007}^{+0.005} $ & $0.043$&${}_{-0.009}^{+0.011} $ & $0.040$&${}_{-0.007}^{+0.008}$ \\ \hline
  -1.005 & $ 0.042$&${}_{-0.007}^{+0.008} $ & $0.026$&${}_{-0.006}^{+0.006} $ & $0.039$&${}_{-0.009}^{+0.010} $ & $0.029$&${}_{-0.004}^{+0.011}$ \\ \hline
  -0.754 & $ 0.066$&${}_{-0.010}^{+0.011} $ & $0.043$&${}_{-0.007}^{+0.007} $ & $0.069$&${}_{-0.011}^{+0.012} $ & $0.072$&${}_{-0.011}^{+0.009}$ \\ \hline
  -0.503 & $ 0.124$&${}_{-0.014}^{+0.016} $ & $0.089$&${}_{-0.012}^{+0.019} $ & $0.130$&${}_{-0.025}^{+0.024} $ & $0.118$&${}_{-0.018}^{+0.013}$ \\ \hline
  -0.251 & $ 0.341$&${}_{-0.052}^{+0.030} $ & $0.325$&${}_{-0.040}^{+0.030} $ & $0.270$&${}_{-0.027}^{+0.037} $ & $0.276$&${}_{-0.024}^{+0.034}$ \\ \hline
   0.000 & $ 0.661$&${}_{-0.050}^{+0.060} $ & $0.819$&${}_{-0.050}^{+0.046} $ & $0.512$&${}_{-0.046}^{+0.046} $ & $0.641$&${}_{-0.030}^{+0.032}$ \\ \hline
   0.251 & $ 0.256$&${}_{-0.025}^{+0.032} $ & $0.303$&${}_{-0.027}^{+0.048} $ & $0.284$&${}_{-0.047}^{+0.033} $ & $0.333$&${}_{-0.034}^{+0.024}$ \\ \hline
   0.503 & $ 0.106$&${}_{-0.014}^{+0.019} $ & $0.103$&${}_{-0.022}^{+0.016} $ & $0.124$&${}_{-0.024}^{+0.023} $ & $0.121$&${}_{-0.014}^{+0.012}$ \\ \hline
   0.754 & $ 0.065$&${}_{-0.012}^{+0.012} $ & $0.041$&${}_{-0.008}^{+0.008} $ & $0.062$&${}_{-0.011}^{+0.011} $ & $0.045$&${}_{-0.006}^{+0.009}$ \\ \hline
   1.005 & $ 0.035$&${}_{-0.008}^{+0.006} $ & $0.021$&${}_{-0.004}^{+0.006} $ & $0.035$&${}_{-0.006}^{+0.010} $ & $0.026$&${}_{-0.004}^{+0.008}$ \\ \hline
   1.257 & $ 0.019$&${}_{-0.006}^{+0.006} $ & $0.025$&${}_{-0.007}^{+0.006} $ & $0.023$&${}_{-0.006}^{+0.005} $ & $0.036$&${}_{-0.007}^{+0.005}$ \\ \hline
   1.508 & $ 0.022$&${}_{-0.007}^{+0.005} $ & $0.021$&${}_{-0.004}^{+0.007} $ & $0.029$&${}_{-0.006}^{+0.007} $ & $0.027$&${}_{-0.004}^{+0.007}$ \\ \hline
   1.759 & $ 0.024$&${}_{-0.005}^{+0.007} $ & $0.037$&${}_{-0.007}^{+0.006} $ & $0.066$&${}_{-0.018}^{+0.016} $ & $0.042$&${}_{-0.006}^{+0.008}$ \\ \hline
   2.011 & $ 0.039$&${}_{-0.007}^{+0.011} $ & $0.046$&${}_{-0.012}^{+0.010} $ & $0.233$&${}_{-0.055}^{+0.034} $ & $0.095$&${}_{-0.015}^{+0.009}$ \\ \hline
   2.262 & $ 0.080$&${}_{-0.013}^{+0.013} $ & $0.106$&${}_{-0.026}^{+0.012} $ & $0.425$&${}_{-0.049}^{+0.076} $ & $0.195$&${}_{-0.026}^{+0.016}$ \\ \hline
   2.513 & $ 0.352$&${}_{-0.048}^{+0.031} $ & $0.258$&${}_{-0.035}^{+0.028} $ & $0.363$&${}_{-0.039}^{+0.064} $ & $0.440$&${}_{-0.027}^{+0.042}$ \\ \hline
   2.765 & $ 0.675$&${}_{-0.041}^{+0.046} $ & $0.439$&${}_{-0.026}^{+0.058} $ & $0.144$&${}_{-0.020}^{+0.020} $ & $0.335$&${}_{-0.033}^{+0.021}$ \\ \hline
   3.016 & $ 0.249$&${}_{-0.027}^{+0.027} $ & $0.248$&${}_{-0.029}^{+0.022} $ & $0.051$&${}_{-0.012}^{+0.010} $ & $0.117$&${}_{-0.010}^{+0.017}$ \\ \hline
\end{tabular}
\caption{Energy-flow distribution in the three-jet plane, normalised to the total invariant mass,  
as a function of the azimuthal angle, $\phi^*$,  in the different regions 
of the ($\xi ,x_1$) plane.}
\label{table4}
\end{center}
\end{table}


\begin{table}
\begin{center}
\begin{tabular} {||c||c|c|c||}      \hline
\multicolumn{4}{|c|}{{\Large{$\frac{d\sigma}{d \eta^*_{\rm jet}}$ (pb)}}}\\ \hline\hline
$d \eta^*_{\rm jet}$ & $\frac{d\sigma}{d \eta^*_{\rm jet}}$ (pb) & (stat.) (pb) & (stat. $\bigoplus$ syst.) (pb)\\ \hline\hline
  [-3.6,-3.0]&    0.650 &$\pm 0.121$&${}_{- 0.185}^{+0.250}$ \\ \hline
  [-3.0,-2.4]&    2.167 &$\pm 0.228$&${}_{- 0.352}^{+0.335}$ \\ \hline
  [-2.4,-1.8]&    4.279 &$\pm 0.333$&${}_{- 0.557}^{+0.972}$ \\ \hline
  [-1.8,-1.2]&    8.193 &$\pm 0.465$&${}_{- 1.034}^{+1.493}$ \\ \hline
  [-1.2,-0.6]&   10.322 &$\pm 0.526$&${}_{- 0.841}^{+1.162}$ \\ \hline
  [-0.6,0.0]&    9.384 &$\pm 0.506$&${}_{- 0.958}^{+0.940}$ \\ \hline
  [ 0.0,0.6]&    7.132 &$\pm 0.429$&${}_{- 0.856}^{+0.704}$ \\ \hline
  [ 0.6,1.2]&    8.359 &$\pm 0.445$&${}_{- 0.837}^{+1.295}$ \\ \hline
  [ 1.2,1.8]&    9.417 &$\pm 0.471$&${}_{- 0.857}^{+1.171}$ \\ \hline
  [ 1.8,2.4]&    9.543 &$\pm 0.605$&${}_{- 1.338}^{+1.241}$ \\ \hline
  [ 2.4,3.0]&    1.039 &$\pm 0.145$&${}_{- 0.203}^{+0.381}$ \\ \hline
\end{tabular}
\caption{Differential cross section for three-jet production
as a function of the jet pseudorapidity,  
$d\sigma/d\eta^*_{\rm jet}$, of the 
three jets.}
\label{table5}
\end{center}
\end{table}


\begin{table}
\begin{center}
\begin{tabular} {||c||c|c|c||}      \hline
\multicolumn{4}{|c|}{{\Large{$\frac{d\sigma}{dP_{\rm T, jet}^* }$ (pb/GeV)}}}\\ \hline\hline
$d P_{\rm T, jet}^*$ (GeV) & $\frac{d\sigma}{d P_{\rm T, jet}^*}$ (pb/GeV) & (stat.) (pb/GeV) & (stat. $\bigoplus$ syst.) (pb/GeV)\\ \hline\hline
 [1.0,3.0]&  3.970 &$\pm 0.182$&${}_{- 0.333}^{+ 0.521}$ \\ \hline
 [3.0,5.0]&  2.291 &$\pm 0.136$&${}_{- 0.244}^{+ 0.378}$ \\ \hline
 [5.0,7.0]&  0.818 &$\pm 0.090$&${}_{- 0.175}^{+ 0.143}$ \\ \hline
 [7.0,9.0]&  0.172 &$\pm 0.039$&${}_{- 0.070}^{+ 0.058}$ \\ \hline
\end{tabular}
\caption{Differential cross section for three-jet production
as a function of the jet transverse 
momentum, $P^*_{\rm T, jet}$, of the most-forward jet.}
\label{table6}
\end{center}
\end{table}


\begin{table}
\begin{center}
\begin{tabular} {||c||c|c|c||c|c|c||}      \hline
\multicolumn{7}{|c|}{{\Large{$\rho(\varphi)$}}}\\ \hline\hline
\multicolumn{1}{|c||}{} & \multicolumn{3}{|c||}{$\rho(\varphi)$ (most-forward jet)} & \multicolumn{3}{|c||}{$\rho(\varphi)$ (most-backward jet)}  \\ \hline\hline
$d \varphi$ (radians) & $\rho(\varphi)$ & (stat.) & (stat. $\bigoplus$ syst.) &  $\rho(\varphi)$ & (stat.) & (stat. $\bigoplus$ syst.) \\ \hline       
 [0.0,0.2]&    1.524&$ \pm  0.057$&${}_{- 0.083}^{+0.086}$ & 2.188 &$\pm  0.063   $&${}_{-0.119}^{+0.084}$\\ \hline
 [0.2,0.4]&    1.500&$ \pm  0.042$&${}_{- 0.061}^{+0.057}$ & 1.273 &$\pm  0.038   $&${}_{-0.054}^{+0.127}$\\ \hline
 [0.4,0.6]&    0.877&$ \pm  0.029$&${}_{- 0.053}^{+0.043}$ & 0.721 &$\pm  0.030   $&${}_{-0.063}^{+0.044}$\\ \hline
 [0.6,0.8]&    0.473&$ \pm  0.020$&${}_{- 0.026}^{+0.053}$ & 0.350 &$\pm  0.017   $&${}_{-0.034}^{+0.024}$\\ \hline
 [0.8,1.0]&    0.300&$ \pm  0.015$&${}_{- 0.019}^{+0.022}$ & 0.219 &$\pm  0.013   $&${}_{-0.023}^{+0.019}$\\ \hline
 [1.0,1.2]&    0.165&$ \pm  0.010$&${}_{- 0.013}^{+0.018}$ & 0.125 &$\pm  0.009   $&${}_{-0.011}^{+0.015}$\\ \hline
 [1.2,1.4]&    0.103&$ \pm  0.008$&${}_{- 0.015}^{+0.010}$ & 0.069 &$\pm  0.006   $&${}_{-0.007}^{+0.010}$\\ \hline
 [1.4,1.6]&    0.047&$ \pm  0.005$&${}_{- 0.008}^{+0.007}$ & 0.025 &$\pm  0.003   $&${}_{-0.005}^{+0.008}$\\ \hline
\end{tabular}
\caption{The differential jet shape, $\rho(\varphi)$, for the most-forward and 
most-backward jets  with $E^*_{\rm jet} >$ 9 GeV in three-jet events, where the 
Pomeron defines the forward direction.}
\label{table7}
\end{center}
\end{table}


\begin{thebibliography}{99}
\bibitem{lrg} ZEUS Collaboration, M. Derrick et al., Nucl. Phys. B 315 (1993) 481;\\
              H1 Collaboration, T. Ahmed et al., Nucl. Phys. B 429 (1994) 477.

\bibitem{dl} A. Donnachie and P.V. Landshoff, Phys. Lett. B 296 (1992) 227.
\bibitem{h1f2d} H1 Collaboration, C. Adloff et al., Z. Phys. C 76 (1997) 613.

\bibitem{zeusf2d} ZEUS Collaboration, M. Derrick et al., Phys. Lett. B 356 (1995) 129;\\
                  ZEUS Collaboration, J. Breitweg et al., Eur. Phys. J. C 5 (1998) 41.
 
\bibitem{ingelman} G. Ingelman and P. Schlein, Phys. Lett. B 152 (1985) 256.

\bibitem{pQCD}   See e.g. {\it Proceedings of the Workshop on Future Physics at HERA}, 
G. Ingelman, A. De Roeck and  R. Klanner (eds.), Vol.~2, DESY, Hamburg, (1996), and references therein.

\bibitem{ryskin} M.G. Ryskin, Sov. J. Nucl. Phy. 52 (1990) 529.
\bibitem{bekw}  N.N. Nikolaev and B.G. Zakharov, Z. Phys. C 5 (1992) 331;\\
                M. W\"usthoff, Phys. Rev. D 56  (1997) 4311;\\
                W. Buchm\"uller, A. Hebecker and M.F. McDermott, Nucl. Phys. B 487 (1997) 283;\\
                J. Bartels et al., Eur. Phys. J. C 7 (1999) 443.

\bibitem{zeusmx} ZEUS Collaboration, J. Breitweg et al., Eur. Phys. J. C 6 (1999) 43.


\bibitem{h1fs} H1 Collaboration, S. Aid et al., Z. Phys. C 70 (1996) 609;\\
               H1 Collaboration, C. Adloff et al., Phys. Lett. B 428 (1998) 206; \\
               H1 Collaboration, C. Adloff et al., Eur. Phys. J. C 1 (1998) 495;\\ 
               H1 Collaboration, C. Adloff et al., Eur. Phys. J. C 6 (1999) 421; \\
               H1 Collaboration, C. Adloff et al., Eur. Phys. J. C 20 (2001) 29. 

\bibitem{zeusfs} ZEUS Collaboration, M. Derrick et al., Phys. Lett. B 332 (1994) 228;\\
                 ZEUS Collaboration, M. Derrick et al.,  Z. Phys. C 67 (1995) 227;\\
                 ZEUS Collaboration, J. Breitweg et al., Phys. Lett. B 421 (1998) 368.

\bibitem{wu} For a review of PETRA results, see e.g.: \\
             S.L. Wu, Phys. Rep. 107 (1984) 59;\\
             B. Naroska, Phys. Rep. 148 (1987) 67.

\bibitem{zeus1} ZEUS Collaboration, M. Derrick et al., Phys. Lett. B 293 (1992) 465;\\
                ZEUS Collaboration, M. Derrick et al., Z. Phys. C 63 (1994) 391.  
\bibitem{zeus2} ZEUS Collaboration, U. Holm (ed.), {\it The ZEUS Detector}, Status Report 
                (unpublished), DESY, 1993, available on \\
                {\tt http://www-zeus.desy.de/bluebook/bluebook.html}.

\bibitem{cal} M. Derrick et al., Nucl. Instr. and Meth. A 309 (1991) 77;\\
              A. Andresen et al.,  Nucl. Instr. and Meth. A 309 (1991) 101;\\
              A. Caldwell et al.,  Nucl. Instr. and Meth. A 321 (1992) 356;\\
              A. Bernstein et al.,  Nucl. Instr. and Meth. A 336 (1993) 23.
\bibitem{ctd} N. Harnew et al., Nucl. Instr. and Meth. A 279 (1989) 290;\\
              B. Foster et al., Nucl. Phys. Proc. Suppl. B 32 (1993) 181;\\
              B. Foster et al., Nucl. Instr. and Meth. A 338 (1994) 254.

\bibitem{fpc} A. Bamberger et al., Nucl. Instr. and Meth.  A 450 (2000) 235.
\bibitem{srtd} A. Bamberger et al., Nucl. Instr. and Meth. A 401 (1997) 63.
\bibitem{lumi} J. Andruszk\'ow et al., DESY 01-041 (2001);\\
               ZEUS Collaboration, M. Derrick et al., Z. Phys. C 63 (1994) 391.
 
 
\bibitem{rapgap} RAPGAP (version 2.08/06), H. Jung, Comp. Phys. Comm. 86 (1995) 147.
\bibitem{cdm} Y. Azimov et al., Phys. Lett. B 165 (1985) 147;\\
  G. Gustafson, Phys. Lett. B 175 (1986) 453;\\
  G. Gustafson  and U. Petersson, Nucl. Phys. B 306 (1988) 746;\\
  B. Andersson, G. Gustafson and L. L\"onnblad, Z. Phys. C 43 (1989) 625.
 
\bibitem{ariadne} L. L\"onnblad, Comp. Phys. Comm. 71 (1992) 15;  Z. Phys. C 65 (1995) 285.

\bibitem{lepto} G. Ingelman, A. Edin and J. Rathsman, Comp. Phys. Comm. 101 (1997) 108.

\bibitem{lund} B. Andersson et al., Phys. Rep. 97 (1983) 31.

\bibitem{jetset} H.-U. Bengtsson and T. Sj\"ostrand, Comp. Phys. Comm. 46 (1987) 43;\\
 T. Sj\"ostrand, Comp. Phys. Comm. 82 (1994) 74.

\bibitem{heracles} K. Kwiatkowski, H. Spiesberger and H.-J. M\"ohring, Comp. Phys. Comm. 69 (1992) 155.

\bibitem{ridi} RIDI (version 2.0), M.G. Ryskin and A. Solano, {\it 
Proceedings of the Workshop on Monte Carlo Generators for HERA Physics}, 
Hamburg, Germany, A.T. Doyle et al., (eds.), DESY (1999), p.~386.


\bibitem{cteq4} H.L. Lai et al., Phys. Rev. D 55 (1997) 1280.
\bibitem{satrap} H. Kowalski, DESY 99-141 (1999).
                   


\bibitem{satura} K. Golec-Biernat and M. W\"usthoff, Phys. Rev. D 59 (1999) 014017;\\
                 K. Golec-Biernat and M. W\"usthoff, Phys. Rev. D 60 (1999) 114023.

\bibitem{taro} T. Yamashita, Ph.D. Thesis, University of Tokyo (2001), in preparation.


\bibitem{jung} H. Jung and H. Kowalski, private communication.

\bibitem{epsoft} M. Kasprzak, Ph.D. Thesis, Warsaw University, DESY F35D-96-16 (1996).

\bibitem{triple} R.D. Field and G. Fox, Nucl. Phys. B 80 (1974) 367;\\
                 A.H. Mueller, Phys. Rev. D 2 (1970) 2963; Phys. Rev. D 4 (1971) 150.

\bibitem{django} K. Charchu{\l}a, G.A. Schuler and H. Spiesberger, Comp. Phys. Comm. 81 (1994) 381.


\bibitem{geant} GEANT 3, R. Brun et al., Technical Report CERN-DD/EE/84-1  (1987).

\bibitem{f2} ZEUS Collaboration, M. Derrick et al., Z. Phys. C 72 (1996) 399.

\bibitem{sinistra} H. Abramowicz, A. Caldwell and R. Sinkus, Nucl. Inst. and Meth. A 365 (1995) 508.

\bibitem{kt} S. Catani et al., Phys. Lett. B 269 (1991) 432.

\bibitem{jade} JADE Collaboration, W. Bartel et al., Phys. Lett. B 91 (1980) 142.
\bibitem{tasso} TASSO Collaboration, R. Brandelik et al., Phys. Lett. B 97 (1980) 453.
\bibitem{x1x2x3}  J. Ellis and I. Karliner, Nucl. Phys. B 148 (1979) 141.


\bibitem{opalshape} OPAL Collaboration, G. Alexander et al., Z. Phys. C 69 (1996) 543.

\end{thebibliography}
\end{document}